\documentclass{sigchi}
\pdfoutput=1

 \toappear{ }

\usepackage{balance}  
\usepackage{graphics} 
\usepackage{txfonts}
\usepackage{mathptmx}
\usepackage[pdftex]{hyperref}
\usepackage{color}
\usepackage{booktabs}
\usepackage{textcomp}
\usepackage{microtype} 
\usepackage{ccicons}  
\usepackage[table]{xcolor}

\usepackage{todonotes}

\def\plaintitle{Towards an Understanding of the Effects of Augmented Reality Games on Disaster Management}

\def\emptyauthor{}

\makeatletter
\def\url@leostyle{%
  \@ifundefined{selectfont}{
    \def\UrlFont{\sf}
  }{
    \def\UrlFont{\small\bf\ttfamily}
  }}
\makeatother
\def\pprw{8.5in}
\def\pprh{11in}

\definecolor{linkColor}{RGB}{6,125,233}
\hypersetup{%
  pdftitle={\plaintitle},
  pdfauthor={\emptyauthor},
  bookmarksnumbered,
  pdfstartview={FitH},
  colorlinks,
  citecolor=black,
  filecolor=black,
  linkcolor=black,
  urlcolor=linkColor,
  breaklinks=true,
}

\begin{document}

\title{\plaintitle}

\numberofauthors{1}
\author{%
  \alignauthor{Markus Luczak-Roesch\\
    \affaddr{School of Information Management}\\
    \affaddr{Victoria University of Wellington}\\
    \affaddr{New Zealand}\\
    \email{markus.luczak-roesch@vuw.ac.nz}
    }
}

\maketitle

\begin{abstract}

Location-based augmented reality games have entered the mainstream with the nearly overnight success of Niantic's \emph{Pok\'emon Go}. Unlike traditional video games, the fact that players of such games carry out actions in the external, physical world to accomplish in-game objectives means that the large-scale adoption of such games motivate people, \emph{en masse}, to do things and go places they would not have otherwise done in unprecedented ways. The social implications of such mass-mobilisation of individual players are, in general, difficult to anticipate or characterise, even for the short-term. In this work, we focus on disaster relief, and the short- and long-term implications that a proliferation of AR games like \emph{Pok\'emon Go}, may have in disaster-prone regions of the world.  We take a distributed cognition approach and focus on one natural disaster-prone region of New Zealand, the city of Wellington. 

\end{abstract}



\section{Introduction}

The nearly overnight success of the game Pok\'emon Go in July 2016 suddenly brought location-based augmented reality (AR) games into our every-day lives. We now see people swiping on their smartphone screens while sitting on the bus, we can read about people getting into dangerous situations because they tried to capture a valuable Pok\'emon\footnote{\url{goo.gl/6mxFXH}}, or we may even play the game ourselves. All of a sudden we are now confronted with a seamless overlap of AR gameplay and the real world, independent from whether we actively take part or just cross the path of players. The implications of this novel intersection aren't very well understood yet, but we can quite certainly expect it to become a central public concern with more games of that kind being released.

Quantitative data about the Pok\'emon Go activities of individuals is -- maybe rightfully so -- not available for research. Furthermore, due to the novelty of the phenomenon, also no qualitative studies have been conducted yet (or at least nothing has been published as of early September $2016$). Hence, we can only investigate this new phenomenon from an abstract viewpoint aiming to crystallize the most relevant hypothesis to inform future research and to raise awareness amongst policy makers so that these are enabled to prepare the adequate policy instruments to deal with this new public concern.

In this research note we report on our ongoing efforts to consolidate information about general effects of location-based AR games on urban disaster management. In particular, we present a case study of a city with high vulnerability to natural disasters -- the city of Wellington in New Zealand -- focused on describing the touching points between the Pok\'emon Go gameplay mechanisms and the city's emergency response plans. Our findings suggest that there is a largely unexplored ambivalence in the promises and perils of location-based AR games with respect to disaster management. This is manifested in an unclear situation about the intersection of public spaces with individuals stimulated by time- and location-dependent incentives of AR games. The implications of this insufficiently investigated area range from uncertainty about the obligations of AR game providers for supporting disaster management not only with mobility data but also by reducing the gameplay interference with crisis locations to an epistemic gap about the behaviour of human AR game players in crisis situations.


\section{Research Method and Setting}



Our research was conducted as a case study involving a single case as it is commonly done in the information systems discipline \cite{benbasat1987case}. Since this is early stage work on a novel research problem, namely the interplay of location-based AR games and urban disaster management, this case study approach is the adequate choice to contribute rigorously informed new hypothesis about the interdependence of artifacts within a complex socio-technical setting \cite{eisenhardt1989building, yin1981case}. This method is anchored within HCI research as kind of \textsl{distributed cognition approach}. We are aiming at "paying close attention to the activities of people and their interactions with material", so far contributing one dimension of a comprehensive ethnographic study, namely the review of documents related to the case, to investigate how "groups of individual agents interacting with each other in a particular environment" \cite{rogers2004new}.

\subsection{Case study setup}
Our case study draws upon the collection of publicly available information from various sources that are officially related to: a) the disaster management and response plans of the city of Wellington; and b) the make up of the Pok\'{e}mon Go gameplay. 

\subsubsection{Wellington: Disaster prone capital city}
Wellington is the capital city of New Zealand, a country that is highly vulnerable to natural disasters of various kinds, mostly due to the fact that it is a group of volcanic islands. Home to about $400,000$ people and a stationary age group distribution (median age of 37.2 years)\footnote{Source: \url{http://goo.gl/pMm21Z}}, Wellington is located at the south end of the north island. Additionally, the city attracts a large amount of tourists and business travellers, counting more than $1.4$ million reported guest nights in various forms of accommodation\footnote{Source: \url{http://goo.gl/TKyF6R}}. Since the midst of the 19th century the city was affected by numerous heavy earhquakes, which has its most likely cause in two fault lines that are located in the wider region. But also heavy weather conditions are a common cause of emergency response in Wellington. Hence, disaster readiness has become a fundamental concern in the every-day life of Wellingtonians, informed and supported by various services provided by the local council as well as the national government.

\subsubsection{Pok\'{e}mon Go: A location-based augmented reality game}
Mid of July 2016 the company Niantec, Inc. released the location-based augmented reality game Pok\'{e}mon Go as a follow up to the game Ingress that falls into the same category. In contrast to Ingress, Pok\'{e}mon Go has become a massive success in its early days featuring a massive burst of player activity. The gameplay narrative in Pok\'{e}mon Go is given as follows: (1) Players create an individually styled avatar; (2) the avatar is placed on a map in the virtual world depending on the location of the player in the real-world; (3) in particular locations players can pick up items with their avatar that feature a Pok\'{e}Stop or they can join team battles with other players' avatars if they are local to a so called Pok\'{e}mon gym; (4) Pok\'{e}mon creatures are algorithmically placed on the map and the players are incentivized for catching them via in-game currencies.

\subsubsection{Investigated documents and material}
By "officially related information" we mean that we only regard information that were published on the Web by organisational entities that act with official capacity related to the two aforementioned two areas. This includes the Wellington City Council, the Ministry of Civil Defence and Emergency Management, the Earthquake and War Damages Commission (EQC), and GNS Science on the one hand as well as Niantic, Inc. on the other. The seed list of URIs for our online research is given in Table\ref{tab:uris}. Starting from this seed list, we thoroughly explored the online presences of the aforementioned information providers and collected all relevant documents including documents provided by other sources but explicitly referred to from these official ones. Relevant documents were investigated in depth and a mapping was created to link a set of core disaster management themes to aspects of the Pok\'{e}mon Go gameplay that may have a specific impact on particular themes.

\begin{table}[htbp]
\centering
\scriptsize
\begin{tabular}{c}
\cellcolor{black!10} \textbf{Seed URIs} \\ \hline \hline
related to disaster management \\ \hline
\url{http://getthru.govt.nz/} \\
\url{http://www.getprepared.org.nz/}\\
\url{http://www.geonet.org.nz/}\\
\url{http://www.gw.govt.nz/}\\
\url{http://www.eqc.govt.nz/}\\ 
\url{http://www.civildefence.govt.nz/}\\
\url{http://wellington.govt.nz/}\\
\url{http://www.nrc.govt.nz/civildefence/}\\
\url{http://www.nzdf.mil.nz/default.htm}\\ \hline
 related to Pokemon Go\\ \hline
\url{http://www.pokemongo.com/} \\
\url{https://support.pokemongo.nianticlabs.com}\\
\end{tabular}
\caption{Seed URIs for our online research, which happened in the time window between $15/07/2016$ and $01/09/2016$.}
\label{tab:uris}
\end{table}
\normalsize

\begin{table*}[ht]
\centering
\scriptsize
\begin{tabular}{p{2cm}|p{1.8cm}|p{1.8cm}|p{1.8cm}|p{4.1cm}|p{4.1cm}}
 \cellcolor{black!10}\textbf{Disaster management themes} & \cellcolor{black!10}\textbf{Places} & \cellcolor{black!10}\textbf{Procedures} & \cellcolor{black!10}\textbf{Stakeholders} & \cellcolor{black!10}\textbf{Location-based AR gameplay mechanisms} & \cellcolor{black!10}\textbf{Hypothetical problematic effects of location-based AR gameplay mechanism on disaster management} \\ \hline \hline 
 \textbf{Disaster awareness} & safe spaces & household emergency plan, practice behaviour & family, care givers & resource demanding interaction design & \textbf{Information dissemination:} People may get into flow within the game and miss important messages passed via other channels about a disaster.\\ \hline 
 & endangered spaces & information dissemination & governmental agencies & combination of location- and time-dependent incentivization &  \textbf{Risk assessment:} People may interpret moderate early warnings in the way that it is still safe enough to approach an endangered location in order to get an advantage over people who already avoid that place. \\ \hline \hline 
 \textbf{Real-time and in situ disaster relief} & critically affected areas, endangered areas, safe spaces & official disaster response actions, information dissemination & governmental agencies, national and international aid organization & combination of location- and time-dependent incentivization &  \textbf{Evacuation (guided or self-managed):} Particular safe spaces get disproportionally and unexpectedly crowded so that not all people can benefit from those safe spaces. \\  \hline \hline 
 \textbf{Post-disaster phases} & critically affected areas, endangered areas, safe spaces & self-relief guidelines, information dissemination & family, care givers, friends & resource demanding interaction design, time-dependent incentivization & \textbf{Self-relief guidelines:} Smartphone batteries may get drained quicker so people are unable to contact relatives and friends to a) inform them about a disaster or b) find out about their situation or position to organize post-disaster help. \\ \hline 
 & critically affected areas, endangered areas, safe spaces & official disaster response actions, information dissemination & governmental agencies, national and international aid organization & combination of location- and time-dependent incentivization & \textbf{Return from evacuation:} People may want to return to the game quickly to avoid being disadvantaged too much. This may result in a tendency to return to areas of a crisis region too early and even before these are declared fully safe. And even when areas are declared safe people coming for gaming may negatively impact the return experience of other affected people (technically in terms of interferences with relief work as well as socially in terms of additional distress). \\ 
\end{tabular}
\caption{Mapping A between disaster management themes and Pok\'{e}mon Go gameplay mechanism: Hypothetical problematic effects.}
\label{tab:map}
\end{table*}
\normalsize

\begin{table*}[ht]
\centering
\scriptsize
\begin{tabular}{p{2cm}|p{1.8cm}|p{1.8cm}|p{1.8cm}|p{4.1cm}|p{4.1cm}}
 \cellcolor{black!10}\textbf{Disaster management themes} & \cellcolor{black!10}\textbf{Places} & \cellcolor{black!10}\textbf{Procedures} & \cellcolor{black!10}\textbf{Stakeholders} & \cellcolor{black!10}\textbf{Location-based AR gameplay mechanisms} & \cellcolor{black!10}\textbf{Hypothetical positive effects of location-based AR gameplay mechanism on disaster management} \\ \hline \hline 
 \textbf{Disaster awareness} & endangered areas, safe spaces & information dissemination & players & resource demanding interaction design, time-dependent incentivization & \textbf{Information dissemination:} A significant change in the behaviour of other players (e.g. disproportionate amount of players disappear from map, team members do not show up for in-game meeting) may trigger awareness of some event in the real world and motivate a player to check by pausing to play for a moment. \\\hline \hline 
 \textbf{Post-disaster phases} & critically affected areas, endangered areas, safe spaces & self-relief guidelines, information dissemination & family, care givers, friends, players, governmental agencies, aid organizations & resource demanding interaction design, time-dependent incentivization & \textbf{Relief (guided or self-managed):} Some people may carry additional equipment (e.g. battery packs, food) in order to reach valuable locations or just generally play longer. This turns them into a particularly interesting group for official relief workers when seeking for solutions to disseminate information or deal with a general resource shortage. \\ 
\end{tabular}
\caption{Mapping B between disaster management themes and Pok\'{e}mon Go gameplay mechanism: Hypothetical positive effects.}
\label{tab:map2}
\end{table*}
\normalsize

\section{Mapping location-based AR gameplay mechanism and disaster management}

In this section we are going to present two mappings between disaster management themes identified in the context of the city of Wellington and a set of generic AR gameplay mechanisms present within Pok\'{e}mon Go. For each mapping we formulate a hypothesis to qualify the potential impact the AR gameplay may have on disaster management in case of an actual disaster. We seek to create an overarching picture of how these disaster management themes can be linked to the generic AR gameplay mechanisms in order to set the line for a constructive embedding of this emerging technology with the particulars of every-day-life in areas that are highly vulnerable to disasters. Hence, Table\ref{tab:map} shows hypotheses about possibly problematic effects in case of a disaster and Table\ref{tab:map2} complements this by showing hypotheses about possibly positive effects.

Our structured review of disaster management material resulted in an agreement that the distinction into \textbf{(1) disaster awareness}, \textbf{(2) real-time and in situ disaster relief}, and \textbf{(3) post-disaster phases} suits best to put the various disaster management activities and recommendations described in the context of the city of Wellington into themes. Most of the official online resources use this pattern to organize information for the citizens independent from the disaster type, which usually is used as the top level categorization. Additionally we find that within these three dimensions three classes of involvements in disaster response occur regularly:\textbf{ (a) places; (b) procedures; (c) stakeholders}.

The review of the selected material describing Pok\'{e}mon Go brought out that the following two gameplay mechanisms stand out most:First, a goal within the game is to capture incentives or compete with others at as many times as possible. But the value of these incentives is not only determined by time (incentives may be available for only a short period of time or may be picked up on a first come first serve basis) but also by location, since incentives captured in different places may have different value. We describe these gameplay mechanisms as \textbf{time-dependent incentivization} and \textbf{location-dependent incentivization} and highlight that in many cases these occur in combination. Second, the incentivizaton strategy is supported by a very extensive UX design addressing multiple senses to get players attention. We summarize this aspect, which is related to the cognitive focus of players as well as the computational load put on their devices, as \textbf{resource demanding interaction design}.


\section{Discussion}

In this section we are going to contextualize our hypothesis generating investigation about the impact of AR games on disaster response with established knowledge from related literature as well as disaster management practice.

\subsection{Can AR game data be used in digital disaster response?}

The emerging field of digital humanitarianism involving digital disaster response has gained wide attention over the recent years \cite{meier2015digital}. Beside the great prospects, which are fueled by widely visible efforts such as Ushahidi and Flowminder, this area is also facing an increasing amount of critical voices \cite{crawford2015limits}. Researchers question the adequacy of using private data without permission but also that, if this data is used, this sometimes happens under non-transparent conditions giving selected parties exclusive access. This does not mean that there is a trivial way out of this situation, because sharing crisis data is an ethically ambivalent dilemma as it always comes with the danger of misleading interpretation if not adequately analysed for example. But while the scholarly world is debating to overcome the transparency and reproducibility crisis in research through data sharing, crisis data seems to be a protected good for a few.

This stress field of digital disaster response is a critical challenge for location-based AR games. While the mobility patterns of players are a great source for quantitative insight before, during, and after a crisis situation, our mapping of gameplay characteristics throws up the following important question: Do people have to be reported when the game data reveals that they were approaching an endangered area to gain points? Is the game provider responsible to take all location-based incentives off the map during a disaster to avoid people being attracted by endangered spaces? Do we need default mechanisms to interrupt attention demanding apps in order to broadcast important crisis-related information? This study cannot give an answer to those question, but it reveals how important it will be to have a critical discussion about them. This is particularly important because \textbf{location-based AR games do not happen in the private spaces of individuals}, they reach out to public spaces and necessarily interfere with public interests in those spaces. All this will become even more important as affordable technologies to make AR increasingly ubiquitous and invasive (e.g. AR goggles and other forms of wearable devices) are on the rise.


\subsection{Do AR games create new barriers in disaster management?}

The barriers to effective crisis relief are a central concern in disaster management research. Typically the scientific inquire is focused on challenges within the disaster management organizations delivering crisis relief. Previous work has identified the typical areas of interest in this regard to be communication and information sharing, authority, and command, as well as coordination and causality flow \cite{o2005abc,bharosa2010challenges}. The actual behaviors of those affected by a disaster has mostly been investigated from the perspective of social resilience and response to crisis \cite{quarantelli1977response, elliott2006race} rather than any real-time and in-situ interdependency with crisis management as in our investigation.

It is noteworthy that this cross-disciplinary viewpoint has long been requested as articulated by Masten and Obradovic in \cite{masten2008disaster}: ``Preparing a large population for any kind of disaster will require a developmental perspective on human resilience, risk, and vulnerability, as well as the integration of ideas on resilience from the sciences of communication, engineering, computing, public health, and ecology, among others.'' And Gunderson emphasizes that "appropriate responses must include anticipating unexpected never-before-experienced effects and impacts" \cite{gunderson2010ecological}. In the light of these demands our study suggests that the \textbf{time- and location-based incentivisation of AR games may change people's usual mobility patterns and impact disaster management due to misleading presuppositions}.

\subsection{Towards socio-technical emergency response augmentation?}

Many countries already provide Emergency Alert Systems (EAS), broadcasting systems that can exploit any available telecommunications, radio, television and satellite infrastructure in order to send out wide ranging or precisely scoped information in case of a disaster situation\footnote{See \url{https://www.fcc.gov/general/emergency-alert-system-eas} and \url{http://www.mass.gov/eopss/agencies/mema/be-prepared/be-informed/} for examples from the US.}. This study raises important new challenges for future EAS. The mapping of problematic effects of AR gameplay mechanisms shows how essential it will be to widen the broadcasting capabilities of EAS to reach players while these are in attention demanding game flow. Furthermore, one can also abstract from the case of AR games and consider any kind of wearable technology or IoT device to become equally important receivers for EAS broadcasting in the future.

Bringing the mapping of positive effects in at this point lets us consider the potential to generally \textbf{augment emergency response with the power of the digitized citizen embodied in decentralised resource hubs made up of wearable technologies and mobile device batteries for example}. While this has the potential to be a promising extension of the existing technical augmentation of response teams \cite{steingart2005augmented}, it also raises important research questions about the psychological and social effects when such a critical responsibility is implicitly passed on to members of the general public.

\section{Conclusion}

Following previous studies on location-based AR games in the context of culture \cite{stark2016playful,chess2014augmented}, health \cite{ma2014virtual}, as well as digital economy\cite{hulsey2014gift}, our work addressed the impact of this emerging phenomenon on disaster management. It raises important questions about the opportunities but also new barriers disaster management is facing due to location-based AR games. The importance of adapting laws and policies to deal with the new phenomenon of location-based AR has already been identified \cite{wassom2014augmented} but needs to be re-emphasized in the light of our suggestions, to ensure that the virtual and the real world can meet in a structured way even in case of unanticipated crisis.

The work described in this note is the tipping point of a larger research agenda. We presented the initial findings of our hypotheses generation process, which was supported by a rigorously designed case study. Even though we can assign to our work that it underwent careful and critical reflection, it does not represent empirical evidence for the potential clashes between urban disaster management and location-based AR games. Because of the vitally important role of disaster management, this needs to be articulated clearly and openly to avoid that any wrong conclusions are drawn that may change disaster management plans under wrong assumptions.

With this work we set the line for subsequent empirical investigation of the actual behavior of location-based AR game players in urban environments and provide policy makers with material to inform their handling of this emerging challenge of public interest. This research agenda, shaped by the outcomes of this initial investigation, aims at running an ethnographic study involving field observations of players of location-based AR games in the Wellington CBD area and focus groups with representatives from different demographic backgrounds. We also seek to expand our structured document analysis to cover material about disaster management planning from further places with a varying vulnerability to a wider range of possible disasters. Ultimately this kind of work will be vital to ensure that augmented reality is not only integrated with the real world for the gameplay and the business purposes of the provider, but also for more effective crisis relief in case of an actual disastrous event.

\balance{}

\bibliographystyle{SIGCHI-Reference-Format}
\bibliography{sample}

\end{document}